\documentclass[twocolumn,prl]{revtex4}
\usepackage{graphicx}
\usepackage{dcolumn}
\usepackage{bm}
\usepackage{hyperref}

\def\>{\rangle}
\def\<{\langle}
\def\u{\!\!\uparrow}

\def\d{\!\!\downarrow}

\def \be{\begin{equation}}
\def \ee{\end{equation}}
\def \beq{\begin{equation}}
\def \eeq{\end{equation}}
\def \bea{\begin{eqnarray}}
\def \eea{\end{eqnarray}}

\begin{document}

\author{Sophia E. Economou$^1$, Netanel Lindner$^{2,3}$ and Terry Rudolph$^4$}
\affiliation{
$^{1}$Naval Research Laboratory, Washington, DC 20375, USA\\
$^{2}$Department of Physics, Technion-Israel Institute of Technology, 32000 Haifa, Israel\\
$^{3}$Institute for Quantum Information, California Institute of Technology, Pasadena, CA 91125, USA\\
$^{4}$Optics Section, Blackett Laboratory, Imperial College London, London SW7 2BZ, United Kingdom}

\title{Optically generated 2-dimensional photonic cluster state from coupled quantum dots}

\date{\today}

\begin{abstract}
We propose a method to generate a two-dimensional cluster state of
polarization encoded photonic qubits from two coupled quantum dot
emitters. We combine the recent proposal \cite{machinegun} for
generating 1-dimensional cluster state strings from a single dot,
with a new proposal for an optically induced conditional phase (CZ)
gate between the two quantum dots. The entanglement between the two
quantum dots translates to entanglement between the two photonic
cluster state strings. Further inter-pair coupling of the quantum
dots using cavities and waveguides can lead to a 2-dimensional
cluster sheet. Detailed analysis of errors indicates that our
proposal is feasible with current technology. Crucially, the emitted
photons need not have identical frequencies, and so there are no
constraints on the resonance energies for the quantum dots, a
standard problem for such sources.
\end{abstract}

\maketitle

Measurement-based quantum computation (MQC) is an alternative to the
well-known `circuit model' of quantum computation
\cite{Rau01}. The main idea in MQC is to robustly
create, upfront, a highly entangled state. Once this `cluster state'
is created, which is the challenging part of this approach, only
single qubit measurements are necessary to perform the actual
computation. In the case of photon polarization qubits, performing
single qubit rotations followed by photon number detection is easily
done with high fidelity, which makes them particularly attractive
for MQC. In fact this is one of the most fault-tolerant
architectures known for quantum computing
\cite{clusterfaulttolerance}, and is particularly tolerant to qubit
losses \cite{Var08}, of importance for optical architectures. The
creation of the initial entangled cluster state is, however, a
difficult problem on which much current research efforts are
focused. To date the most promising methods have involve optical
interference of nearly identical photons
\cite{fusion}. By contrast, our proposal here
allows for direct generation of the entangled photons.

In Ref. \cite{machinegun} a proposal was developed for generating a
linear (one-dimensional) cluster state of polarization encoded
photons from single photon emitters with a certain energy level
structure, such as those found in quantum dots (QDs). The relevant
states of the QD are the two spin states $|\u\>,|\d\>$ of the
electron along the optical axis $z$ and the two optically excited
states called trions, which have total angular momentum $3/2$ and
have spin projections along the $z$-direction of $\pm 3/2$ - states
we denote $|3/2\>,|\bar 3/2\>$. The broken symmetry of the QD along
the $z$ axis sets a  preferred direction, along which the optical
polarization selection rules are circularly polarized, and
energetically separates the excited trion states with total angular
momentum $\pm 1/2$ (the light hole states) from these heavy-hole
trion states. In the process of linear cluster state generation
\cite{machinegun} the heavy hole trions are the only excited states
that are populated. The main idea in \cite{machinegun} is to shine a
periodic train of optical linearly polarized $\pi$ pulses, to an
electron that is in a superposition state $|\u\rangle+|\d\rangle$,
exciting it to a superposition of the two trion states
$|3/2\rangle+|\bar 3/2\rangle$. Because QDs have large dipole
moments, spontaneous emission is very fast, both compared to atoms
and to the other relevant time scales in the QD dynamics, at least
for very low magnetic fields. Therefore the trion will spontaneously
decay to the electron state almost instantaneously upon excitation,
emitting a photon of either right ($R$) or left ($L$) circular
polarization, thereby effecting transitions
$|3/2\>\rightarrow|\u\>|R\>,|\bar 3/2\>\rightarrow|\d\>|L\>$. The
state of the emitted photon+spin is $|\u\rangle |R\rangle +
|\d\rangle |L \rangle$ - i.e. they are entangled as both
recombination paths take place simultaneously. The remaining degrees
of freedom of the system are the same, so they are factored out and
omitted for brevity. Subsequent precession of $\pi/2$ radians by the
spin about a weak magnetic field oriented in the $y-$direction is
performed, denoted $R_y(\pi/2)$, before subjecting the dot to
another pulse excitation+emission process. Repeating this protocol
results in a one-dimensional entangled chains of photons.
Importantly, errors were shown to localize and not affect the whole
chain.

Here we will develop an explicit, all-optical protocol for
generating a two-dimensional cluster state comprised by linking two
linear chains like the ones of \cite{machinegun} by controlled phase
(CZ) gates. To do so we present a new proposal, related to that of
\cite{economoucz}, for performing an optically-controlled CZ gate
between two quantum dots. Taking advantage of the exchange
interactions between electrons and the hole, this gate actually
proves to be faster than that of \cite{economoucz} and so is of
independent interest. Crucially, unlike the scheme of
\cite{economoucz}, this process is also compatible with the
operation of the single-dot photonic machine guns, as it performs
the optical CZ in the $z$ basis. The entangled emitters therefore
generate photons which are themselves entangled. Explicitly, an
entangled chain can be created. This circumvents the need for
`fusion gates' \cite{fusion}. Moreover, in our approach, the
photons need not be identical in frequency, so that there are no
constraints on the resonance energies of the two QDs.

\begin{figure}
\includegraphics[width=9.0cm,clip=, bb=2.5cm 6.5cm 19.5cm 17cm]{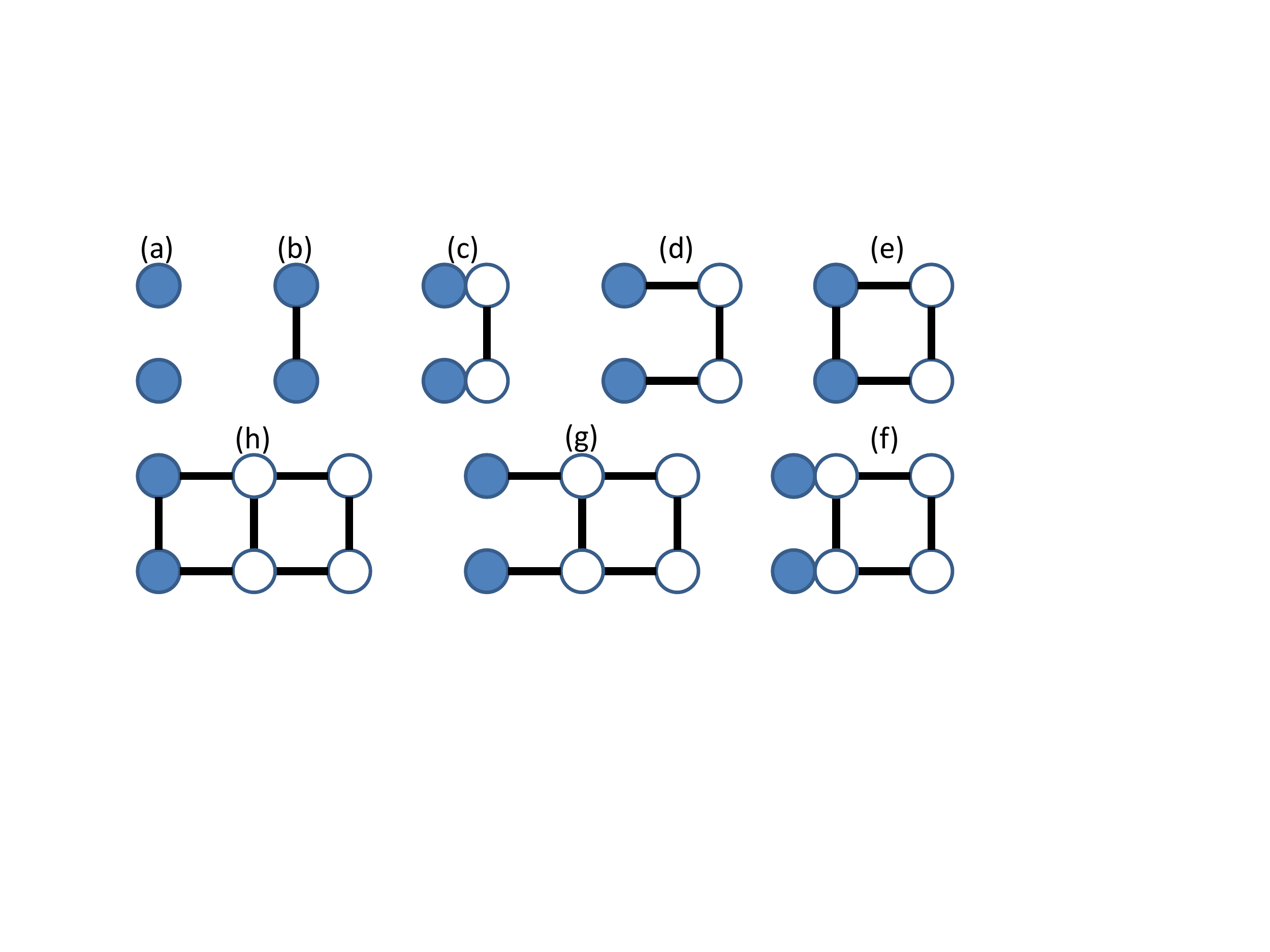}
\caption{A sequence of diagrams depicting the generation of the
cluster state using the standard diagrammatic representations of
such states. The electronic qubits are depicted as filled circles,
the initial electronic state is taken to be $|\u\>|\u\>$. At step
(a) both spins precess under $R_y(\pi/2)$, at (b) the interdot $CZ$
gate is applied, at (c) a pulse excitation followed by trion decay
produces photons (open circles). These procedures are then repeated,
leading to the states of (d),...(h). Details of the states produced
are in the text. Note that to recover the standard form of cluster
states one must use a mapping where the logical qubit $|1\>$ state
is equivalent to the photonic state $-|L\>$, this is because, for
practical reasons, $R_y(\pi/2)$ gates are used instead of Hadamard
gates.} \label{cluster}
\end{figure}

The state evolution for the idealized abstract protocol is depicted
in Fig.~1, for a quantum circuit logically equivalent to the
protocol see Fig.~2. For simplicity we assume that the two QDs are
initialized in the spin up state $|\u\>|\u\>$ (in fact no
initialization is necessary - it can be effected later via
measurements on the photons). First we apply a $R_y(\pi/2)$
operation on each spin yielding $(|\u\>+|\d\>)(|\u\>+|\d\>)$, as in
Fig.~1(a). This is followed by a \textsc{C-Z} gate entangling the
dots, $(|\u\>|\u\>+|\u\>|\d\>+|\d\>|\u\>-|\d\>|\d\>)$, producing the
bond in Fig.~1(b). Immediately after this we apply the pump pulse to
each dot, and the creation of the subsequent photons yield the state
$(|\u\>|R\>|\u\>|R\>+|\u\>|R\>|\d\>|L\>+|\d\>|L\>|\u\>|R\>-|\d\>|L\>|\d\>|L\>)$.
In the circuit of Fig.~2 this is equivalent to the \textsc{CNOT}
gates. This resulting state is equivalent to a 2-qubit cluster
state, where the logical state $|0\>$ ($|1\>$) is redundantly
encoded \cite{fusion} in 2 qubits as $|\u\>|R\>$ ($-|\d\>|L\>$).
Graphically such a situation is depicted with the circles for each
qubit adjacent to each other, Fig.1(c).  A second $R_y(\pi/2)$ on
each dot pushes out the redundantly encoded qubits (i.e., creates a
bond between them in the cluster state), Fig.1~(d), and we start the
cycle anew.

The inter-dot \textsc{C-Z} gate is implemented optically by coupling
to trion states which are higher in energy than the ones used for
the single dot photon emission. These higher-energy trion states are
delocalized, i.e. the voltage bias is such that one of the electrons
in these higher energy states is tunnel-coupled, in contrast to the
single-electron ground states (denoted $|B\>$ and $|T\>$ for bottom
and top QD respectively) and the lower energy trion states, which
are isolated from one another and localized to their respective
quantum dots. This trion mediating the inter-dot interaction has two
electrons in the $|B\>$ and $|T\>$ (``s''-type) states, and the
third electron in the first excited (``p''-type) orbital, which we
take to be the one that is a delocalized (also called `molecular' or
`extended') state, denoted $|E\>$. It has been shown experimentally
that such a regime is feasible \cite{stinaff}. The hole is taken to
occupy a single orbital state $|H\>$, which for simplicity we take
to be completely confined to one QD (this assumption does not
deleteriously affect the overall proposal).

With the three electrons in distinct orbital states, the spin
configuration can acquire any of its allowed values by adding the
three angular momenta. So, for a given orbital configuration there
are a total of eight electron states (two $S=1/2$ doublets and one
$S=3/2$ quadruplet) and two hole spin states, making a total of
sixteen states \cite{economoucz}. By tuning the laser appropriately,
some of the states can be safely ignored due to the large e-e
exchange splittings (on the order of 5 meV) compared to the laser
bandwidth.
\begin{figure}\label{circuit}
\includegraphics[width=6.5cm,clip=, bb=3.5cm 0.1cm 19.5cm 18.7cm]{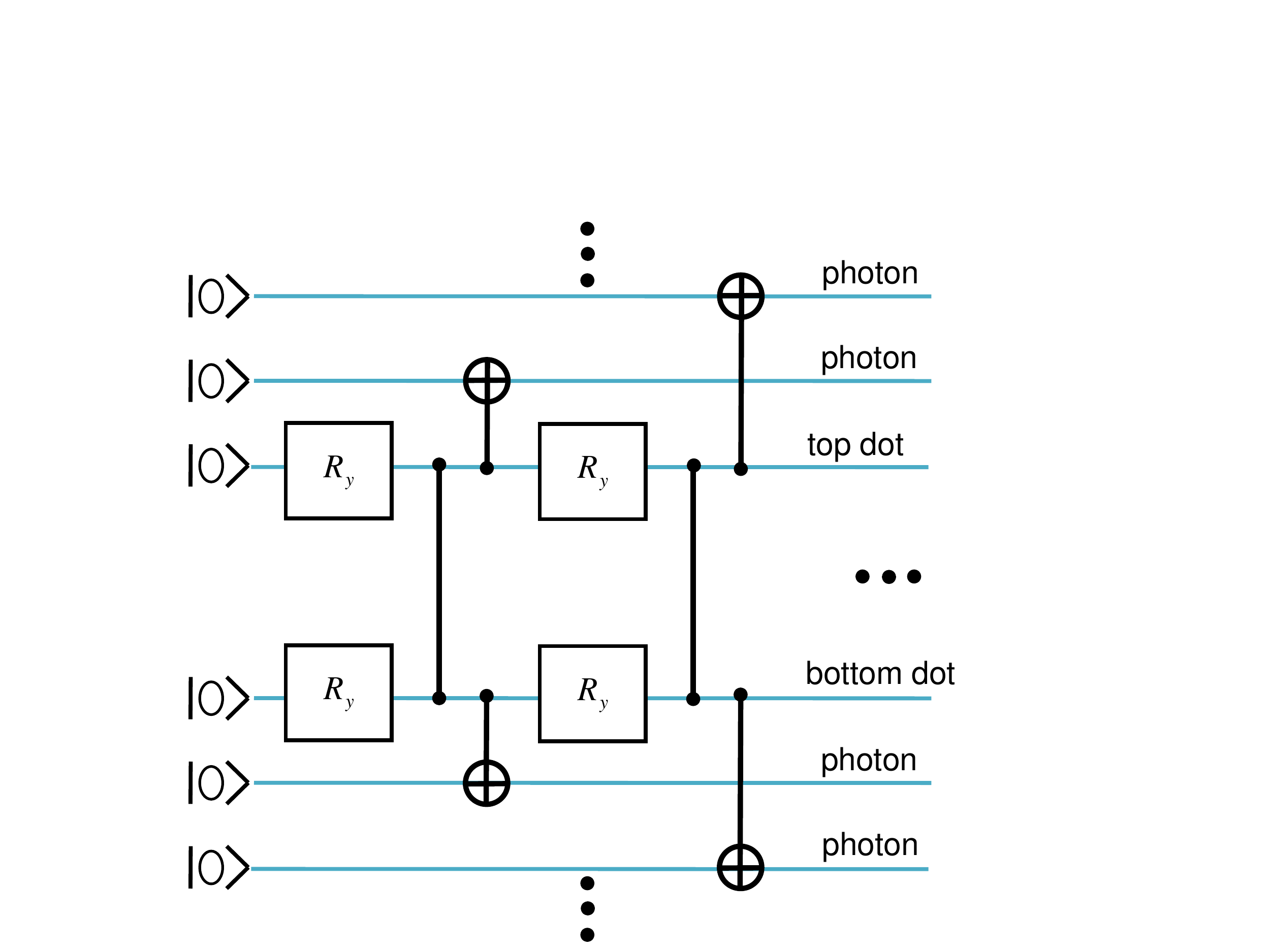}
\caption{A quantum circuit which is logically equivalent to the
idealized evolution of the two QDs. The CZ gates correspond to the
interdot coupling, the CNOT gates to photon emission, and the $R_y$
to precession by $\pi/2$ around a magnetic field in the $y$
direction.} \label{circuit}
\end{figure}

We henceforth focus on the $S=3/2$ states, which have a separable (product) form
of orbital and spin states:
\begin{eqnarray}
|3/2\rangle &=& |A\rangle |\uparrow\uparrow\uparrow\rangle  \label{th} \\
|1/2\rangle &=& |A\rangle \left(|\uparrow\uparrow\downarrow\rangle +
|\uparrow\downarrow\uparrow\rangle + |\downarrow\uparrow\uparrow\rangle
\right)/\sqrt{3} \\
|\bar{1}/2\rangle &=& |A\rangle
\left(|\downarrow\downarrow\uparrow\rangle +
|\downarrow\uparrow\downarrow\rangle + |\uparrow\downarrow\downarrow\rangle
\right)/\sqrt{3} \\
|\bar{3}/2\rangle &=& |A\rangle |\downarrow\downarrow\downarrow\rangle
\label{mth}
\end{eqnarray}
where
\begin{eqnarray}
|A\rangle &=& \frac{1}{\sqrt{6}}\big( |TBE\rangle - |BTE\rangle -
|TEB\rangle \nonumber
\\
&+& |BET\rangle - |EBT\rangle + |ETB\rangle \big).
\end{eqnarray}

The electron and hole in semiconductors are coupled by exchange
interactions. In QDs, these are quite strong (on the order of, or
stronger, than typical Zeeman energies), and they are separable into
`isotropic' and `anisotropic' terms \cite{Bayer_Finestructure}. The
isotropic term is much stronger - typical values of this are 0.3-0.5
meV - so it is the leading term in our parameter regime. Its
physical origin is the lack of inversion symmetry (along the growth
direction) in the QD. We will ignore the anisotropic term, which
originates from in-plane asymmetry (deviation of the QD cross
section from a disk) and is typically small, in the order of $\mu$eV
\cite{Bayer_Finestructure,scheibner}; its effects can be incorporated as standard errors in the
gate.

The Hamiltonian is therefore given by $
\mathcal{H}=\sum_i\alpha_z(r_i,r_h) s_{iz} j_z, $ where
$\alpha_z(r_i,r_h)$ is an operator acting on the envelope
wavefunctions of the electrons and the hole. The index $i$ runs over
the three electrons, $r_h$ denotes the position of the hole, and $z$
is the growth axis. The operator $j$ acts on the hole spin, which we
take to be a pseudospin, $j_z=\pm 3/2$ (i.e., ignore light hole
states). Adding and subtracting terms, we can rewrite $\mathcal{H}$
as
\begin{eqnarray}
\mathcal{H}&=&\frac{1}{3}\sum_i\alpha_z(r_i,r_h) S_{z} j_z \label{ham}\\
 &+& \frac{1}{3}\!\!\!\!\sum_{(ij)\in\atop \{(12),(23),(31)\}}\!\!\!\!(\alpha_z(r_i,r_h)-\alpha_z(r_j,r_h)) (s_{zi}-s_{zj})j_z,  \nonumber
\end{eqnarray}
with $S_z=\sum_i s_{iz}$. The first term on the RHS of Eq. (\ref{ham}) conserves the total electron
spin, so it only has nonzero matrix elements within the three total spin
subspaces discussed above. The second set of terms has nonzero matrix
elements only between different total electron spin states. Since typical values of
the electron-electron exchange are about one order of magnitude more than
typical electron-hole exchange interactions, we can ignore the total spin
mixing terms and focus on the Hamiltonian
\begin{eqnarray}
\mathcal{H}\simeq\frac{1}{3}\sum_i\alpha_z(r_i,r_h) S_{z} j_z ,  \label{hamap}
\end{eqnarray}
and only consider the states (\ref{th})-(\ref{mth}) tensored with the hole
state, which is an $8\times 8$ space. The mean value of
the operator $\alpha$ in state $|A\rangle|H\rangle$ is
\begin{eqnarray}
&&\sum_i\langle H|\langle A| \alpha(r_i,r_h) |H\rangle |A\rangle \nonumber\\
&=& \frac{2\times 3}{6}\sum_{K=B,T,E}
\langle H|\langle K| \alpha(r,r_h) |H\rangle |K\rangle  \nonumber\\
&\equiv & \delta_0^{BH} + \delta_0^{TH} + \delta_0^{EH}.
\end{eqnarray}

Assuming that the hole is localized in one of the two quantum dots, say the
one labeled by $B$, we have $\delta_0^{TH}=0$. We will define the sum of the
nonzero terms to be $\delta_0$. Now we have the operator
\begin{eqnarray}
\mathcal{H}_{3/2}=\frac{\delta_0}{3} S_z j_z,
\end{eqnarray}
acting only on the spin states. Clearly, this operator is already diagonal
in the basis we have chosen. Since it is invariant under the simultaneous
flip of $S_z$ and $j_z$, we expect the states to be doubly degenerate. Then the eigenenergies and
corresponding eigenstates are:
\begin{eqnarray}
E_1=\frac{\delta_0}{4} \text{ with eigenstates } |3/2\rangle |\Uparrow\rangle,
|\bar{3}/2\rangle |\Downarrow\rangle \\
E_2=\frac{{\delta_0}}{12} \text{ with eigenstates } |1/2\rangle
|\Uparrow\rangle, |\bar{1}/2\rangle |\Downarrow\rangle \\
E_3=-\frac{{\delta_0}}{12} \text{ with eigenstates } |1/2\rangle
|\Downarrow\rangle, |\bar{1}/2\rangle |\Uparrow\rangle \\
E_4=-\frac{\delta_0}{4} \text{ with eigenstates } |3/2\rangle
|\Downarrow\rangle, |\bar{3}/2\rangle |\Uparrow\rangle
\end{eqnarray}
The states with energy $E_1$ are dark. The remaining ones are optically
accessible. We are particularly interested in the states with energy $E_4$. These
states are coupled only to the two-qubit states $|\uparrow\uparrow\rangle$
and $|\downarrow\downarrow\rangle$ by polarization $\sigma^-$ and $\sigma^+$
respectively. The two-qubit states $|\uparrow\downarrow\rangle$
and $|\downarrow\uparrow\rangle$ couple to
the states with $E_2, E_3$ with these polarizations. We take
advantage of the energy splitting between $E_4$ and $E_2, E_3$ to selectively address only the two-qubit $%
|\downarrow\downarrow\rangle$ state and realize the \textsc{C-Z} gate.

For simplicity we fix the polarization of the pulse to $\sigma^+$ (behaviour for the orthogonal
polarization is found by flipping all the spins). If we label the dipole matrix
element for transition $|\downarrow\downarrow\rangle \rightarrow |\bar{3}%
/2\rangle |\Uparrow\rangle$ to be $d_0$. Then only the triplet state
$|T_+\>$ couples to the excited state $
 |\bar{1}/2\rangle |\Uparrow\rangle
$ with dipole strength  $ \sqrt{\frac{2}{3}} d_0$.

Given these three transitions, we can implement the \textsc{c-z}
gate by acting with a resonant 2$\pi$ pulse on the
$|\downarrow\downarrow\rangle$ state and avoid coupling to the other
transitions.

We now turn to a consideration of the various sources of errors and
imperfections. A crucial feature of our proposal is the fact that
all non-leakage errors in the system \emph{localize}. By non-leakage
errors we refer to any decoherence which eventually returns the
electrons back into the computational subspace - ie back into any
state such that one electron is located in the orbital ground state
of each dot. By localize we refer to the fact that the action of any
decoherence map on the electrons is (mathematically) equivalent to a
(different) decoherence map on some of the emitted photons, however
crucially the number of affected photons is at most the four photons
emitted around the time the decoherence event occurs. This ensures
that the final output state takes the form of an ideal cluster
subject to localized random noise--a noise model for which fault
tolerant procedures are known to work. In particular we emphasize
that this allows for production of photonic cluster states for
arbitrarily longer times than the electron decoherence timescales
might suggest.

The error localization might be seen in quite a general manner as
follows. Consider the quantum circuit of Fig.~\ref{circuit} encoding
the generic evolution. Let some decoherence occur which is described
by a set of Kraus operators $\{K_i\}$ acting on the spin only. If we
denote by $U$ the unitary evolution which corresponds in the figure
to the circuit consisting of four photon emissions (i.e., two
photons per dot and including the CZ gate acting between the dots)
then an error and subsequent evolution takes the generic form
\[
\rho_i'=U(I\otimes I\otimes K_i)(\rho_{spin}\otimes |0\>\<0|\otimes |0\>\<0|)(I\otimes I\otimes K_i^\dagger)U^\dagger.
\]
It is a remarkably nice feature of this process that in fact we can find a Kraus operator $\tilde{K_i}$ acting now only on the four  emitted photons, such that
\[
\rho_i'=( \tilde{K_i}\otimes I)U(\rho_{spin}\otimes |0\>\<0|\otimes |0\>\<0|)U^\dagger(I\otimes \tilde{K_i}^\dagger).
\]
Physically this means that an error occuring on the spin is
(mathematically) identical to some different error occurring on the
photons subsequently emitted. Crucially it affects only the next
four photons, and no more--hence the term localization.

We now discuss some specific sources of error, and their expected impact.

\textsl{1. Imperfect CZ gate}---If we label $\Omega _{0}$ the Rabi
frequency of the target transition from $|\downarrow\downarrow\rangle$, then the other transitions see a Rabi frequency of $%
\Omega _{1}=\Omega _{0}/\sqrt{3}$ and $\Omega _{2}=\sqrt{\frac{3}{2}}\Omega
_{0}$, with a large detuning. As such some population is transferred to those excited states and
it is not returned via stimulated emission. Instead, the incoherent process of spontaneous emission
redistributes that population. For simplicity we assume that the small population transferred is equal for
the two unwanted transitions and that spontaneous emission equally redistributes it. The simplest way to express the Kraus operators
$\{K_j\}$ describing the generalized quantum evolution in the two spin qubit subspace is by one nearly unitary, CZ operator:
\[
K_0=u_1|\uparrow \uparrow\rangle\langle\uparrow \uparrow|
+|\psi^-\rangle\langle\psi^-|+u_2|\psi^+\rangle\<\psi^+|
-|\downarrow \downarrow \rangle\<\downarrow\downarrow|,
\]
plus eight more operators
describing the redistribution of the populations.
For a pulse of a total duration of ~40 ps and for anisotropic exchange $\delta=0.5$ meV we have $%
|u_{1}|\simeq |u_{2}|\sim 0.99$. Then the remaining operators,
$\{K_1,K_2,...,K_8\}$, are $\frac{\sqrt{1-|u_1|^2}}{2}
|k\rangle\langle 1|$ and $\frac{\sqrt{1-|u_2|^2}}{2}
|k\rangle\langle 3|$, with $k=1,2,3,4$. Since the operator sum
representation is not unique, we can find a different set of Kraus
operators $\{M_j\}$ for which $M_0$ is proportional to the
\textsc{CZ} gate. Setting $u_1=u_2\equiv u$, these are $M_0 = \alpha
~\text{CZ}, M_1 = e^{i\phi}\sqrt{1-\alpha^2} K_0
-\frac{\alpha}{\sqrt{2}} \left(K_1+K_2\right)$, and
$M_2=\frac{1}{\sqrt{2}}\left(K_1-K_2 \right)$, with
$\phi=\arctan\left({\frac{Im(u)}{1-Re(u)}}\right)$. For $j=3,...,8$,
$M_j=K_j$. The value of $\alpha$ is a measure of how close the
operation is to a unitary \textsc{CZ}. For $u_1=u_2=0.99$ we find
$\alpha=0.98$, with an error of $1-\alpha=0.02$. Physically we can
therefore interpret the action of the gate as follows: With
probability $\alpha^2$ we obtain a perfect CZ gate, with probability
$(1-\alpha^2)$ we obtain some other type of evolution.

\textsl{2. Unequal g factors}---In general, the two QDs comprising
the QD molecule will have different g factors, and therefore
different precession frequencies. This means that we cannot get both
spins to undergo a $R_y(\pi/2)$ operation solely based on
precession. One can correct for this mismatch by spin-echo type
control by applying to the fast spin at time
$\tau=\pi(\omega_f^{-1}-\omega_s^{-1})/4$ a single qubit $\pi$
rotation about the optical axis to delay it ($\omega_f$, $\omega_s$
are the fast and slow Zeeman splittings respectively). These
rotations are by design fast (in the ps regime) \cite{economouprb06}
and have been demonstrated experimentally \cite{greilichrot}.

\textsl{3. Decay of one of the two resident electrons into the other
QD}---When this error occurs it will cause both quantum dots to stop
emitting photons, and thus amounts to a detectable loss error on the
cluster state. We want the hole to occupy the dot for which the
single particle energy is lowest so that recombination will now
correct for this error. In general we expect the computation to be
quite resilient to such loss. Depending on the height of the
tunneling barrier between the dots, eventually the system will decay
back into the desired computational basis with one electron in each
dot.

\textsl{4. Precession during CZ gate}---This is an error whose
effect will be essentially the same as the case of the single dot
machine gun\cite{machinegun}. It results in a localizable error,
which, provided the magnetic field strength is chosen suitably, can
be extremely low.

In conclusion, we have developed a scheme for generation of $2\times
N$ dimensional photonic cluster state based on coupled quantum dots.
Analysis of the relevant errors shows our proposal to be robust and
feasible with current state of the art systems. This scheme can be
generalized to the generation of a two-dimensional sheet either by
considering multiple stacked dots, or by employing cavities and
waveguides to couple distant dots. Future work will include specific
cavity-waveguide-quantum dot designs for generation of a cluster
state sheet of arbitrary dimensions.

This work was supported in part by the Engineering and Physical Sciences Research Council and the US Office of Naval Research.

\end{document}